\documentclass[pre,floatfix,twocolumn,showpacs]{revtex4}
\usepackage{graphicx}
\usepackage{bm}
\usepackage{amsmath}
\usepackage{amssymb}
\def\ArrhenFitD{0.195}
\def\ArrhenFitB{0.307}
\def\ArrhenFitC{0.318}
\begin{document}
\preprint{APS/123-QED}
\title{Intermittent relaxation in hierarchical energy landscapes}
\author{A. Fischer}
\email{andreas.fischer@physik.tu-chemnitz.de} \affiliation{Institut f\"ur
Physik, Technische Universit\"at Chemnitz, D-09107 Chemnitz, Germany}
\author{P. Sibani}
\email{paolo.sibani@ifk.sdu.dk} \affiliation{Institut for Fysik og Kemi, SDU,
DK-5230 Odense M, Denmark}
\author{K. H. Hoffmann}
\email{hoffmann@physik.tu-chemnitz.de} \affiliation{Institut f\"ur Physik,
Technische Universit\"at Chemnitz, D-09107 Chemnitz, Germany}
\begin{abstract}
We numerically simulate a thermalization process in an energy landscape with
hierarchically organized metastable states. The initial configuration is
chosen to have a large energy excess, relative to the thermal equilibrium
value at the running temperature. We show that the initial energy surplus is
dissipated in a series of intermittent bursts, or quakes, whose rate decreases
as the inverse of the age of the system. In addition, one observes energy
fluctuations with a zero centered Gaussian distribution. These pertain to the
pseudo equilibrium dynamics within a single metastable state, and do not
contribute to the energy dissipation. The derivative of the thermal energy
with respect to the logarithm of time is asymptotically constant, and
comprises a temperature independent part, and a part with an Arrhenius
temperature dependence. The findings closely mirror recent numerical
simulation results obtained for microscopic glassy models. For these models,
record-sized energy fluctuations have been claimed to trigger intermittent
events during low temperature thermalization. In the present model
record-sized fluctuations are by construction needed to trigger changes from
one metastable state to another. This property thus suffices to explain the
statistical property of intermittent energy flow in complex metastable
systems.
\end{abstract}
\maketitle
\section{Introduction}
Many characteristics of low temperature glassy dynamics are only weakly
related to details of the microscopic interactions. Aging processes, for
example, generically follow a change of an external parameter, e.g. a
temperature quench. In experimental glassy systems a sequence of large,
so-called intermittent, configurational re-arrangements can be observed, which
generate non-Gaussian tails in the Probability Density Function (PDF) of
configurational probes
\cite{bissig.h.03.intermittent.21,%
buisson.l.03.intermittency.s1163,%
kegel.w.00.direct.290,%
weeks.e.00.three-dimensional.627,%
cipelletti.l.03.time-resolved.s257,%
buisson.l.03.intermittent.603}. %
In microscopic model systems
\cite{sibani.p.05.intermittency.563,%
sibani.p.06.mesoscopic.69}, %
the PDF of the energy fluctuations following a thermal quench also features a
zero centered Gaussian and an exponential tail which covers large negative
changes. The former describes pseudo-equilibrium fluctuations and the latter
is related to shifts between different metastable attractors. These and other
features have been recently analyzed using an Edwards-Anderson spin glass
model
\cite{sibani.p.05.intermittency.563}, %
and an even simpler Ising model with four spin plaquette interactions
\cite{sibani.p.06.aging.031115}, %
which is known to possess central features of glassiness, e.g. a metastable
super-cooled phase and an aging phase
\cite{lipowski.a.00.cooling-rate.6375,%
swift.r.00.glassy.11494}. %
For both models, the rate of intermittent energy flow out of the system was
found to fall off with the reciprocal of the system age. For the plaquette
model, the temperature dependence of the rate was also analyzed in detail.

The above aging properties can be understood using the idea that the
attractors dynamically selected during the process are marginally stable
\cite{sibani.p.93.slow.1482,%
sibani.p.03.log-poisson.8}. %
The so-called \emph{record dynamics} scenario
\cite{sibani.p.03.log-poisson.8,%
sibani.p.05.intermittency.563} %
then links the intermittent events, or \emph{quakes}, to record-sized energy
fluctuations occurring within thermalized local domains. Record fluctuations
are not associated to a definite scale. For this reason alone, they will not
lead to observable effects, unless the energy landscape supporting the
fluctuations is self-similar under a change of scale. Conversely, within a
self-similar energy landscape, record-sized fluctuations are required to
induce attractor changes.

Simple hierarchical models of configuration space are already known to explain
many facets of complex relaxation
\cite{vincent.e.91.slow.209,%
bouchaud.j.95.aging.265,%
krawczyk.j.02.low-temperature.302,%
sibani.p.89.hierarchical.2853,%
sibani.p.93.emergent.479,%
hoffmann.k.97.age.613}. %
In these mesoscopic descriptions, the configuration space of a physical system
is coarse grained into a graph, whose nodes represent lumped sets of
microscopic configurations with similar energies. Each node is thus simply
characterized by its energy, i.e. the typical energy of its constituents, and
by a degeneracy, i.e. the number of its constituents. Lumping is physically
reasonable if the microscopic configurations lumped into the same node are
able to reach a state of local thermal equilibrium on a time scale short
compared to the time it takes to access configurations belonging to other
nodes. Reference~\cite{sibani.p.93.emergent.479} provides an example where the
lumping is explicitly carried out starting with a microscopic model.
Connections between different nodes represent possible dynamical pathways. In
tree models, the connectivity is at the lowest level possible for a connected
configuration space. The unique path between two arbitrary nodes represents
the dominant path in the original dynamical problem, i.e., typically for
thermal dynamics, the path through the lowest possible energy barrier. In the
tree graph shown in figure~\ref{fig:Setup:Tree} the vertical axis represents
the energy, and a hierarchy of barriers of different sizes is present. By
construction larger and larger barriers must be overcome starting from one of
the bottom nodes in order to access larger sets of states. Correspondingly,
any sub-tree contains a number of lesser sub-trees characterized by smaller
barriers. Also note that the energy minima have different energies and
represent physically different meta-stable configurations. This inequivalence
was introduced in the context of the so-called LS tree model
\cite{sibani.p.91.relaxation.423,%
hoffmann.k.97.age.613} %
of which figure~\ref{fig:Setup:Tree} shows a generalization. The LS model's
scale invariance is restricted to a discrete set of energy re-scalings due to
the presence of only two `elementary' energy scales, i.e. the energy
differences $L$ and $S$ between neighboring nodes. To approach full re-scaling
symmetry, the present study considers a randomized version of the model. While
exact analytical results are then no longer possible, the randomized model
still offers a strikingly simple and general conceptualization of a highly
complex relaxation behavior. In particular, only record-sized energy
fluctuations can induce transitions out of a sub-tree into a larger subtree
containing states of lower energy. Comparing the statistics of intermittent
energy flow in a hierarchically structured energy landscape with the
corresponding properties of microscopic models provides a direct check of
whether record-sized energy fluctuations indeed are the main mechanism for
thermal relaxation in complex systems.

In this article, energy traces are obtained from isothermal simulations of an
ensemble of randomized LS model aging after a thermal quench. The energy
fluctuations are treated as time-resolved calorimetry data. Their probability
density function (PDF) shows the fingerprints of intermittent heat flow. The
form of the local density of states of the attractors is extracted from the
Gaussian part of the fluctuations, and the rate of energy flow is extracted
from the tail. The information is analyzed mathematically and related to the
known geometrical properties of the model.

\section{Modified LS-tree ensembles}
In spatially extended systems with short range interactions, one expects
thermalization to occur independently within a number of slowly growing
domains or clusters of neighboring degrees of freedom. The configuration space
of the full systems correspondingly factorizes into a product of configuration
spaces, each term belonging to a single domain. Domains are expected to have a
hierarchical organization, which, in the present context, is rendered by the
modified LS tree described below.

\begin{figure}
{\includegraphics[width=1.00\linewidth]{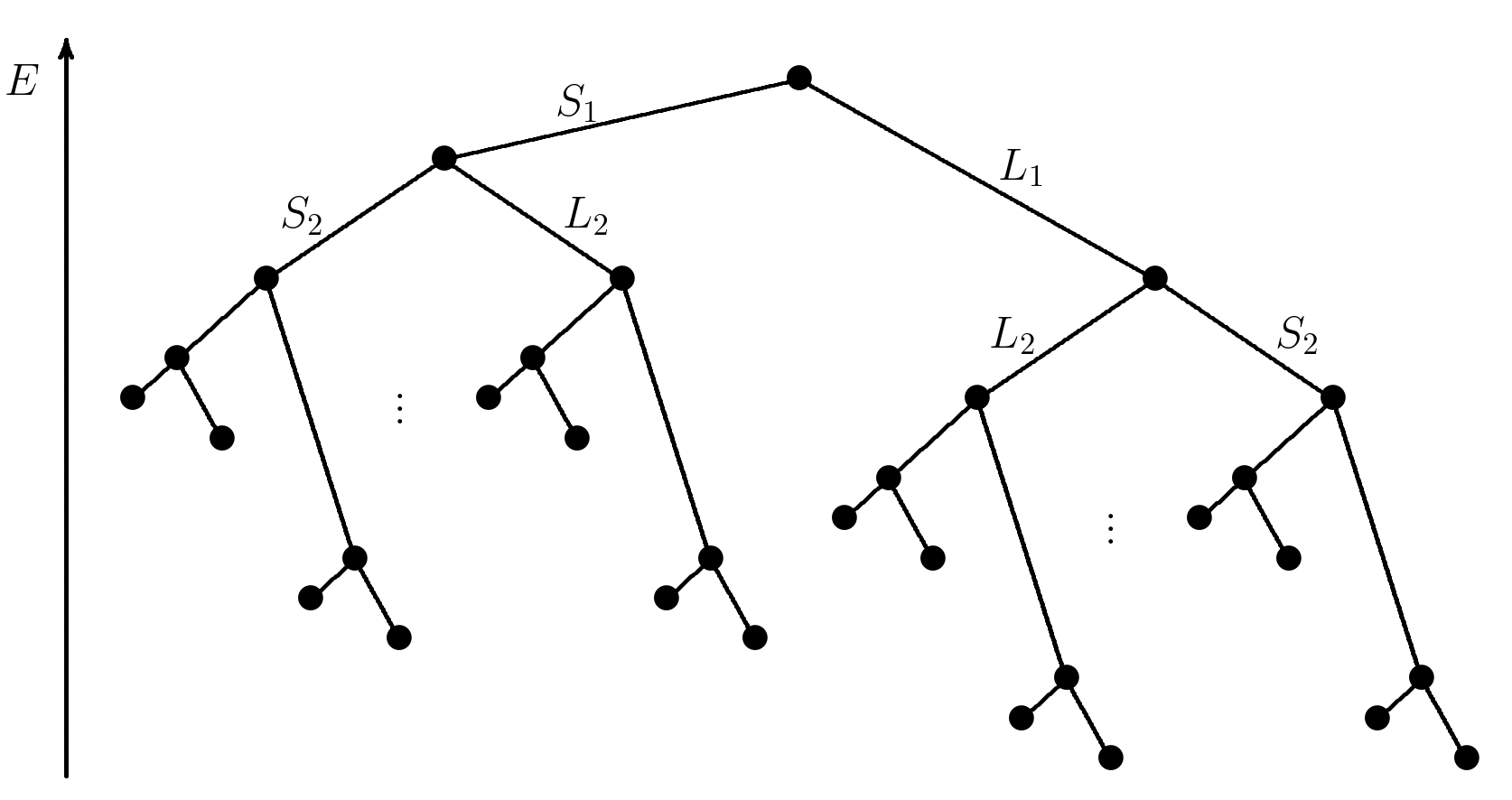}}%
\caption{The randomized LS tree used as model energy landscape in the
simulations. Contrary to previous LS tree setups the $L$ and $S$ branches are
not of equal length for the whole tree but level dependent. This extension
leads to a diversification of the time scales present within the tree and thus
breaks unphysical logarithmic oscillations in the energy decay.}
\label{fig:Setup:Tree}%
\end{figure}
The randomized LS model used in the simulations can be pictured as an upward
rooted tree with the vertical scale representing the energy as shown in
figure~\ref{fig:Setup:Tree}. The $n$'th level of the tree comprises all the
nodes connected to the root by precisely $n$ edges, with $n=0,\ldots,N$. With
the exception of the 'bottom' level with index $n=N$, each node is connected
to two `daughters' of lower energy by a `Longer' and a `Shorter' edge. The
modification of the original model
\cite{sibani.p.91.relaxation.423,%
hoffmann.k.97.age.613} %
consists in choosing, independently for each level, new random values of $L$
and $S$, as discussed further below. Accordingly, the energy differences along
the `Longer' and `Shorter' edges are $\Delta{}E=L_n$ and $\Delta{}E=S_n$ for
each level $n$. The above procedure removes the oscillation on a logarithmic
timescale characteristic of a regular tree, concomitantly decreasing the
energy scale corresponding to the smallest barrier in the system. Each node
represents a lumped set of configurations and hence possesses a degeneracy. In
the model, the degeneracy of a node equals the sum of the degeneracies if its
$L_n$ and $S_n$ daughters, each multiplied by a factor $\kappa_{_{L_n}}$ and
$\kappa_{_{S_n}}$, respectively. Bottom nodes are not degenerate. With this
prescription, the overall degeneracy of the model increases in a nearly
exponential fashion, as the level index increases from the bottom. A list of
values
\begin{equation}
\mathcal{E}=\left\{2^{\frac{i}{4}-1}\;:\;i=0\ldots19\right\}
\;\;\text{.}%
\label{eqn:EED:EnergieListe}%
\end{equation}
describes the possible energy differences. For each level of the tree, two
values are drawn independently and with uniform probability from the list.
They may be equal, and are assigned to the respective $L$ and $S$ branches,
ensuring that $L\ge{}S$.

Thermal relaxation is modeled as hopping between neighboring nodes, with up
and down rates defined by:
\begin{equation}
\Gamma_{{\rm up},j} = f_j \kappa_j {\rm e}^{-\beta \Delta E_j} \qquad
\Gamma_{{\rm down}, j} = f_j \enspace,
\label{eqn:tree:rates}%
\end{equation}
where the index $j\in \{L_n,S_n\}$ labels $L$- and $S$-edges at level $n$. The
hopping rates obey the detailed balance condition
$\Gamma(x,y)P_{\mathrm{eq}}(y)=\Gamma(y,x)P_{\rm eq}(x)$, where the
equilibrium distribution is the Boltzmann distribution. So-called kinetic
factors $f_j$ are introduced to control the relaxation rate along each edge.
Factors independent of $j$ entail a trivial rescaling of the (arbitrary) time
unit. In contrast, non-uniform factors considerably modify the dynamics on
short time scales after a thermal quench
\cite{uhlig.c.95.relaxation.409}. %
E.g. choosing $f_{S_n}\gg{}f_{L_n}$ favors downward transitions along the $S$
edges, ensuring that an initial quench from the top node preferably ends near
the shallowest metastable minimum. We achieve the same in a more ad hoc
fashion, by choosing the highest lying and most shallow local energy minimum
as the initial state for the aging dynamics.

\section{The dynamics}
The dynamical evolution of a single tree is studied using a rejectionless
continuous time Monte Carlo Method
\cite{dall.j.01.faster.260}, %
which operates with an intrinsic time variable $t$, initially set to zero. The
following three steps are iterated: for the current node, the possible
transitions according to (\ref{eqn:tree:rates}) are considered. In general one
upward and two downward transitions are available, except for the bottom
states and the top state, which have no downward, respectively upward
transitions. Among the neighbors to the current node, one is chosen with
probability equal to the transition rate along the corresponding  edge,
divided by the sum of all rates out of the current node. A random waiting time
is then drawn from an exponential distribution with average equal to the
reciprocal of the transition rate. This waiting time is then added to the
global time, and the position of the walker is finally updated.

Each tree in the ensemble is updated independently as described above. The
result is an ensemble of time series, each describing the history of energy
fluctuations of one member of the ensemble. The time series describing the
total energy of the ensemble is then obtained by interweaving the time series
pertaining to each tree into a single data stream ordered by increasing values
of the time variable.

In the following, the symbol $t$ stands for the system age, i.e. the time
elapsed since the beginning of the simulation. The symbol $t_\mathrm{w}$ is
the age of the system when data collection for the fluctuation PDF begins.
Energy fluctuations are energy differences over a time interval $\delta t \ll
t_\mathrm{w}$.
\section{Energy Transfer Statistics}
Our results are based on an ensemble of $2000$ trees of height $N=12$, with
$5000$ independent runs on each tree. The model is set up with degeneracy
growth parameters $\kappa_{_L}=\kappa_{_S}=2.20$ and kinetic factors
$f_L=0.25$ and $f_S=1.00$. The latter of these values would generally carry
units. However, the choice of the units does not affect the results so they
can be considered as being arbitrary here as well as in the figures showing
the numerical results. The data pertain to systems initially quenched into the
shallowest of the available minima, i.e. the minimum connected to the top node
by a series of $S$ links.

\begin{figure}
{\includegraphics[width=1.00\linewidth]{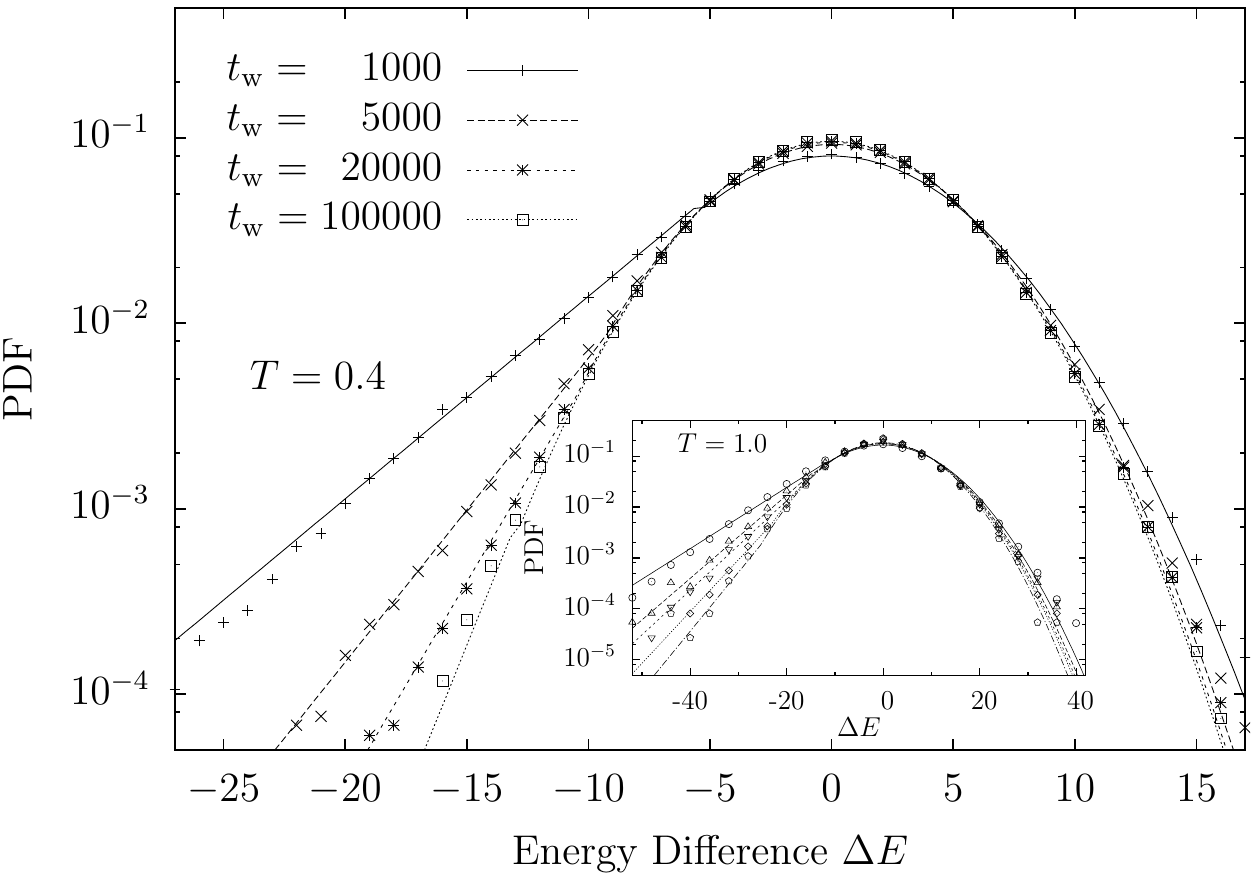}}%
\caption{The PDF of the amounts of energy exchanged between system and thermal
bath within the time $\delta{}t=100$. Negative values represent an energy
outflow, while positive values represent an energy inflow. The data are taken
for four logarithmic equidistant times in the interval $t=[10^3\ldots10^6]$.
The inset (taken from
\cite{sibani.p.06.aging.031115}) %
shows an analogous figure obtained for simulation of a microscopic glassy
model with plaquette interaction.}
\label{fig:EED:SchulterZeit}%
\end{figure}

The PDF of the amount of energy exchanged within a time interval $\delta t$ at
system age $t_\mathrm{w}+k\,\delta{}t$
\begin{equation}
\Delta{}E=E(t_\mathrm{w}+k\,\delta t)-E(t_\mathrm{w}+(k-1)\,\delta t)
\label{eqn:EED:WaermeAustausch}%
\end{equation}
is collected over $k=1\ldots100$ time intervals using $\delta{}t=1$ and
starting at different values of $t_\mathrm{w}$, in the range $10^3 \ldots
10^5$. The zero centered Gaussian peak, flanked on the left by an exponential
tail seen in figure~\ref{fig:EED:SchulterZeit} is the characteristic signature
of intermittency. When $k$ and $\delta t$ are kept fixed, the relative weight
of the intermittent tail decreases with increasing $t_\mathrm{w}$. The width
of the Gaussian fluctuations is independent of $k$ and $\delta t$. We can then
conclude that the fluctuations are (pseudo) equilibrium energy fluctuations
within partially equilibrated subtrees. Clearly then, as the Gaussian
fluctuations do not contribute to the net energy flow, the latter is carried
exclusively by the tail of the distribution, which pertains to intermittent
quakes.

The decay of the average energy is shown in figure~\ref{fig:EED:Energie}:
\begin{figure}
{\includegraphics[width=1.00\linewidth]{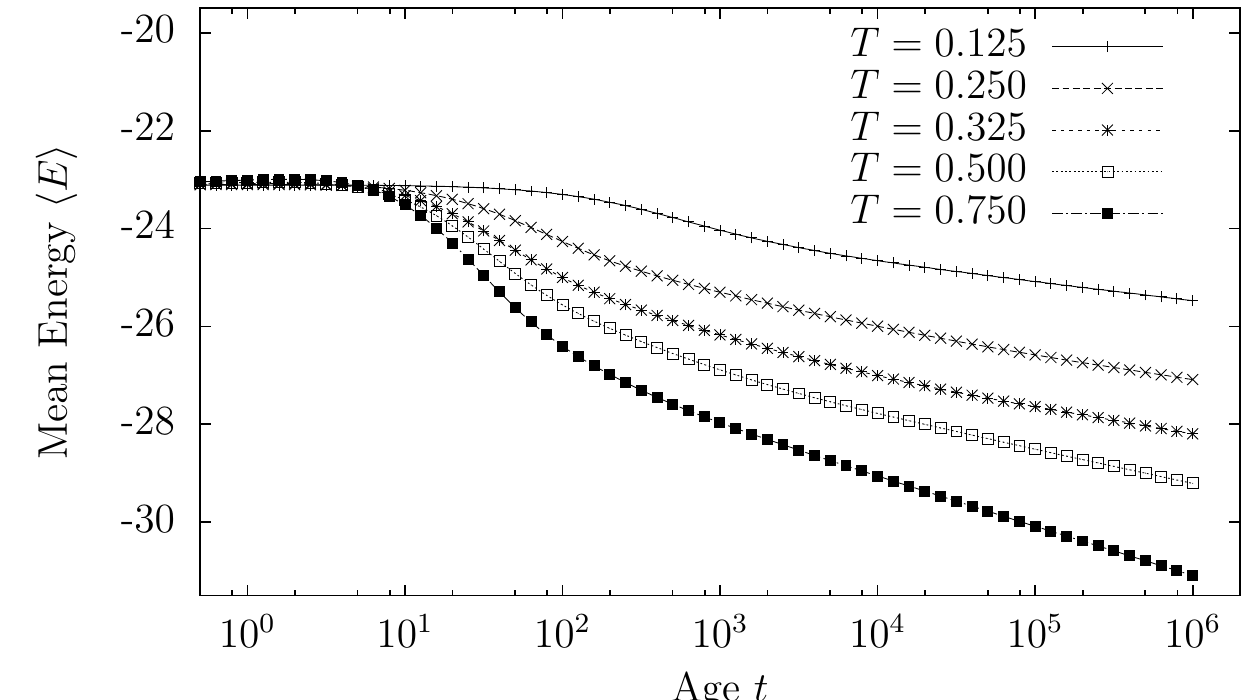}}%
\caption{The mean inner Energy $E$ of the tree ensemble plotted versus system
age for different temperatures showing the progress of the relaxation process.
Note that after some kind of startup phase lasting until $t\approx200$ the
energy shows a practically straight decrease at logarithmic time scale.}
\label{fig:EED:Energie}%
\end{figure}
After an initial transient stretching to approximately $t = 200$, the energy
decreases in time in a logarithmic fashion:
\begin{equation}
E(t,T)=-a(T)\,\ln(t)+\mathcal{C}
\;\;\text{.}%
\label{eqn:EED:EnergieZeit}%
\end{equation}
In this model, a quake occurs when the lowest value of the energy seen in the
simulation decreases. This event corresponds to a first visit to a new subtree
containing states of lower energy, and is -- by construction -- triggered by a
record-sized (positive) energy fluctuation. The logarithmic rate of energy
loss $a(T)$ is thus the product of the logarithmic rate at which record-sized
energy fluctuations occur with the average net amount of energy given off in a
single quake. The logarithmic rate of record-sized energy fluctuation is
independent of both temperature and time
\cite{sibani.p.03.log-poisson.8,%
krug.j.07.records.P07001}. %
The average amount of energy given off in a single quake would also be
temperature independent if the path leading to the new lowest energy state
were purely exothermal. However, since side branches in the tree can trap the
walker on his way down, an activated contribution can be expected.
Accordingly, we expect the thermal energy reached at a fixed time to be higher
the lower the temperature, which is of course a hall mark of the
non-equilibrium nature of the dynamics. The temperature dependent offset
$\mathcal{C}$ is related to the initial stage of the relaxation and has no
importance for our treatment.

According to (\ref{eqn:EED:EnergieZeit}) the rate of energy loss
$r_E=-\mathrm{d}E/\mathrm{d}t$ has the form
\begin{equation}
r_E(t,T)=-\frac{\mathrm{d}E}{\mathrm{d}t}=\frac{a(T)}{t}
\;\;\text{,}%
\label{eqn:EED:EnergieRate}%
\end{equation}
which fully agrees with the numerical results shown in
\begin{figure}
{\includegraphics[width=1.00\linewidth]{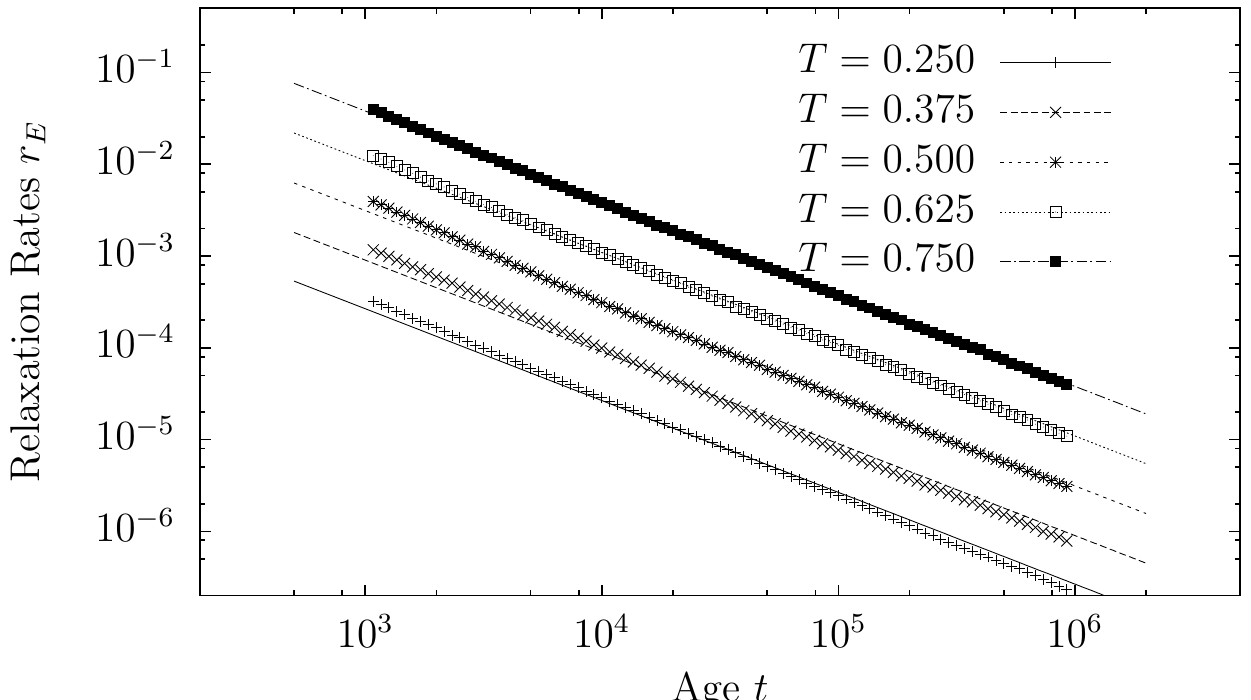}}%
\caption{The rate of energy loss $r_E$ is plotted versus system age for
different temperatures. The lines have the form $r_E=a(T)/t$ with the constant
$a$ as in equations~(\ref{eqn:EED:EnergieZeit})
and~(\ref{eqn:EED:EnergieRate}). To avoid clutter, the data set for each $T$
is multiplied by $1$, $3$, $9$, $\ldots$ in order of increasing $T$. A similar
but much smaller shift of the curves in the same direction is present due to
the temperature dependence of $a$.}
\label{fig:EED:Raten}%
\end{figure}
figure~\ref{fig:EED:Raten} as well as with similar results for microscopic
model simulations
\cite{sibani.p.06.aging.031115,%
sibani.p.05.intermittency.563}. %
The figure depicts the rate of energy loss for selected temperatures (symbols)
together with fits according to (\ref{eqn:EED:EnergieRate}) (lines).
\begin{figure}
{\includegraphics[width=1.00\linewidth]{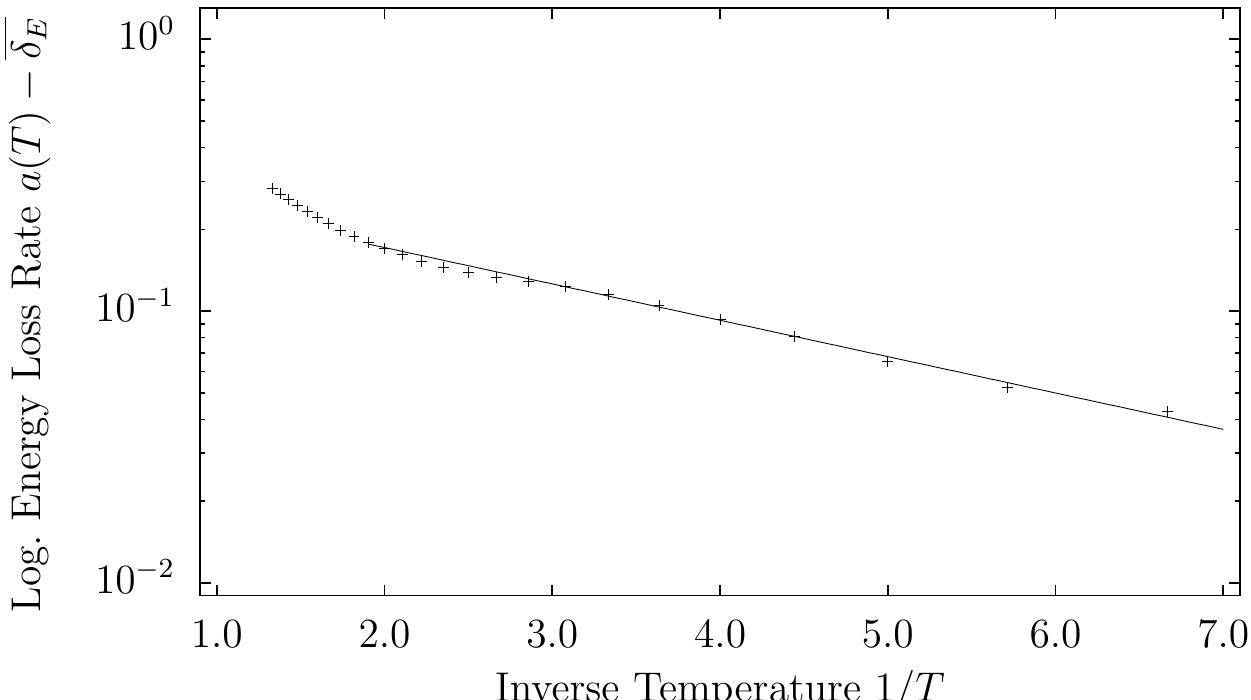}}%
\caption{The logarithmic rate of energy loss $a(T)$ (see
(\ref{eqn:EED:EnergieRate})), shifted by $\overline{\delta_E}$, is plotted vs.
the inverse temperature. The data are fitted to the expression
$a(T)=\overline{\delta_E}+c\exp(-b/T)$ yielding
$\overline{\delta_E}=\ArrhenFitD$, $b=\ArrhenFitB$ and $c=\ArrhenFitC$.}
\label{fig:EED:Arrhenius}%
\end{figure}

As just discussed, we expect the logarithmic rate of energy loss, $a(T)$ to
have the form
\begin{equation}
a(T)= \overline{\delta_E} + c\exp(-b/T)
\;\;\text{,}%
\label{eqn:EED:LogRateForm}%
\end{equation}
where $\overline{\delta_E}$ is the average contribution to the energy given
off by paths not involving any energy barriers. The activated term involves a
barrier which should be comparable in size to the smallest barrier present in
the system. Figure~\ref{fig:EED:Arrhenius} confirms
equation~(\ref{eqn:EED:LogRateForm}) for the present model. The value
$\overline{\delta_E} = \ArrhenFitD$ is the one judged to provide the best
numerical agreement with the data. The parameters $b$ and $c$ where estimated
by least square fitting. The value of the `barrier' parameter $b=\ArrhenFitB$
is close to the value of the smallest energy barrier present in the model,
$b_\mathrm{min}=0.5$.

\begin{figure}
{\includegraphics[width=1.00\linewidth]{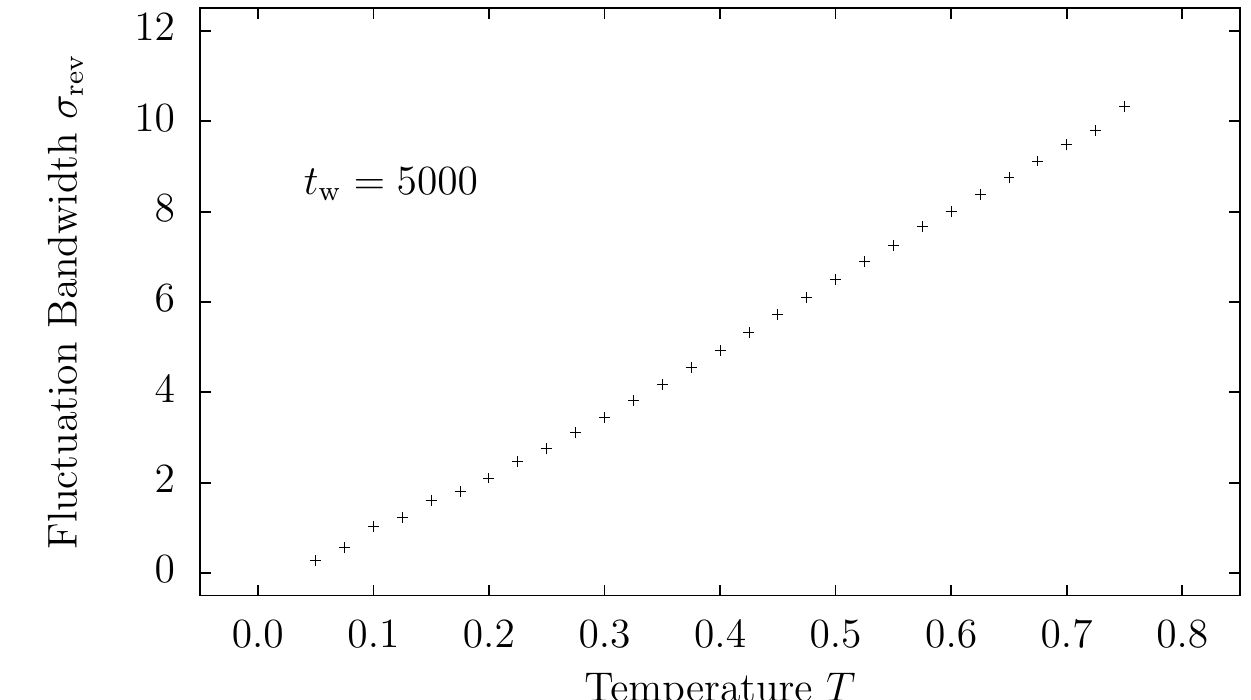}}%
\caption{The standard deviation $\sigma_\text{rev}$ plotted versus temperature
is obtained from the Gaussian part of the energy exchange PDF.}
\label{fig:EED:Reversible}%
\end{figure}
Figure~\ref{fig:EED:Reversible} shows the bandwidth of the Gaussian
fluctuation $\sigma_\text{rev}$ versus the temperature. The curve is
relatively featureless and increases with temperature as expected. For a
classical system (with constant heat capacity) the fluctuation bandwidth would
be proportional to $T$. The data shown have a slightly faster than linear
increase with $T$.

The results presently obtained, with the possible exception of
figure~\ref{fig:EED:Reversible}, very closely mirror the behavior of the Ising
model with plaquette interactions recently investigated by one of the authors
\cite{sibani.p.06.aging.031115}. %

\section{Discussion and conclusion}
Thermal hopping on a tree structure provides a simple framework to describe
relaxation phenomena in complex systems
\cite{hoffmann.k.85.random.401,%
hoffmann.k.88.diffusion.4261,%
sibani.p.89.hierarchical.2853,%
sibani.p.91.relaxation.423,%
ball.k.96.from.963,%
geppert.u.97.hierarchical.l393,%
hoffmann.k.99.slow.30,%
oliveira.v.99.metastable.8793,%
hordijk.w.03.shapes.3671,%
wales.d.03.energy.book,%
schubert.s.04.aging.118,%
schubert.s.06.structure.191}. %
The hierarchical model studied in this work is a modification of the even
simpler LS model
\cite{sibani.p.91.relaxation.423}, %
whose two elementary energy scales are replaced with a wider spectrum of
energy scales. This is physically more realistic, and removes unphysical
logarithmic oscillations
\cite{schreckenberg.m.85.long.483} %
in the energy decay.

The intermittent energy relaxation of this model is strikingly similar to that
observed at microscopic models
\cite{sibani.p.06.aging.031115,%
sibani.p.05.intermittency.563}. %
For the model, the behavior can be understood as follows: At any given time,
the probability distribution is, to a good approximation, supported within a
finite subtree. A larger subtree must be entered in order to lower the
hitherto lowest energy. Thus, record-sized energy fluctuations trigger, with a
certain probability, quakes which then lead to the attainment of lower
energies. Since the records occur at a rate proportional to $1/t$
\cite{sibani.p.93.slow.1482,%
sibani.p.03.log-poisson.8,%
krug.j.07.records.P07001}, %
the rate of quakes is also proportional to $1/t$ and the quantity
$r_E\times{}t=-a(T)$ is, modulo a constant proportionality factor, the energy
given off in a single quake.

Further evidence for the crucial role of record-sized fluctuation was provided
by an analysis of the Edwards-Anderson spin-glass, which was done using two
different dynamical update rules: thermal hopping
\cite{dall.j.03.exploring.233} %
and extremal optimization
\cite{boettcher.s.05.comparing.317}. %
In both cases, it was found that a record-sized energy barrier needed to be
overcome in order to lower the current lowest energy seen so far in the
simulation. Since the update rules are completely different in the two cases,
and yet lead to the same result, the connection between record-sized energy
barriers and low energy states is likely to be a true geometrical property of
complex energy landscapes.

In a thermalizing hierarchical model, record-sized energy fluctuations are
needed to move from one metastable region of configuration space to a
different region of lower energy. In this paper we have shown that this one
property suffices to generate realistic spectra for the intermittent energy
fluctuations.
\bibliography{article}

\begin{thebibliography}{36}
\expandafter\ifx\csname natexlab\endcsname\relax\def\natexlab#1{#1}\fi
\expandafter\ifx\csname bibnamefont\endcsname\relax
  \def\bibnamefont#1{#1}\fi
\expandafter\ifx\csname bibfnamefont\endcsname\relax
  \def\bibfnamefont#1{#1}\fi
\expandafter\ifx\csname citenamefont\endcsname\relax
  \def\citenamefont#1{#1}\fi
\expandafter\ifx\csname url\endcsname\relax
  \def\url#1{\texttt{#1}}\fi
\expandafter\ifx\csname urlprefix\endcsname\relax\def\urlprefix{URL }\fi
\providecommand{\bibinfo}[2]{#2}
\providecommand{\eprint}[2][]{\url{#2}}

\bibitem[{\citenamefont{Bissig et~al.}(2003)\citenamefont{Bissig, Romer,
  Cipelletti, Trappe, and Schurtenberger}}]{bissig.h.03.intermittent.21}
\bibinfo{author}{\bibfnamefont{H.}~\bibnamefont{Bissig}},
  \bibinfo{author}{\bibfnamefont{S.}~\bibnamefont{Romer}},
  \bibinfo{author}{\bibfnamefont{L.}~\bibnamefont{Cipelletti}},
  \bibinfo{author}{\bibfnamefont{V.}~\bibnamefont{Trappe}}, \bibnamefont{and}
  \bibinfo{author}{\bibfnamefont{P.}~\bibnamefont{Schurtenberger}},
  \bibinfo{journal}{PhysChemComm} \textbf{\bibinfo{volume}{6}},
  \bibinfo{pages}{21} (\bibinfo{year}{2003}).

\bibitem[{\citenamefont{Kegel and van Blaaderen}(2000)}]{kegel.w.00.direct.290}
\bibinfo{author}{\bibfnamefont{W.~K.} \bibnamefont{Kegel}} \bibnamefont{and}
  \bibinfo{author}{\bibfnamefont{A.}~\bibnamefont{van Blaaderen}},
  \bibinfo{journal}{Science} \textbf{\bibinfo{volume}{287}},
  \bibinfo{pages}{290} (\bibinfo{year}{2000}).

\bibitem[{\citenamefont{Weeks et~al.}(2000)\citenamefont{Weeks, Crocker,
  Levitt, Schofield, and Weitz}}]{weeks.e.00.three-dimensional.627}
\bibinfo{author}{\bibfnamefont{E.~R.} \bibnamefont{Weeks}},
  \bibinfo{author}{\bibfnamefont{J.~C.} \bibnamefont{Crocker}},
  \bibinfo{author}{\bibfnamefont{A.~C.} \bibnamefont{Levitt}},
  \bibinfo{author}{\bibfnamefont{A.}~\bibnamefont{Schofield}},
  \bibnamefont{and} \bibinfo{author}{\bibfnamefont{D.~A.} \bibnamefont{Weitz}},
  \bibinfo{journal}{Science} \textbf{\bibinfo{volume}{287}},
  \bibinfo{pages}{627} (\bibinfo{year}{2000}).

\bibitem[{\citenamefont{Buisson
  et~al.}(2003{\natexlab{a}})\citenamefont{Buisson, Ciliberto, and
  Garcimart\'\i{}n}}]{buisson.l.03.intermittent.603}
\bibinfo{author}{\bibfnamefont{L.}~\bibnamefont{Buisson}},
  \bibinfo{author}{\bibfnamefont{S.}~\bibnamefont{Ciliberto}},
  \bibnamefont{and}
  \bibinfo{author}{\bibfnamefont{A.}~\bibnamefont{Garcimart\'\i{}n}},
  \bibinfo{journal}{Europhys. Lett.} \textbf{\bibinfo{volume}{63}},
  \bibinfo{pages}{603} (\bibinfo{year}{2003}{\natexlab{a}}).

\bibitem[{\citenamefont{Buisson
  et~al.}(2003{\natexlab{b}})\citenamefont{Buisson, Bellon, and
  Ciliberto}}]{buisson.l.03.intermittency.s1163}
\bibinfo{author}{\bibfnamefont{L.}~\bibnamefont{Buisson}},
  \bibinfo{author}{\bibfnamefont{L.}~\bibnamefont{Bellon}}, \bibnamefont{and}
  \bibinfo{author}{\bibfnamefont{S.}~\bibnamefont{Ciliberto}},
  \bibinfo{journal}{J. Phys.: Condens. Matter} \textbf{\bibinfo{volume}{15}},
  \bibinfo{pages}{S1163} (\bibinfo{year}{2003}{\natexlab{b}}).

\bibitem[{\citenamefont{Cipelletti et~al.}(2003)\citenamefont{Cipelletti,
  Bissig, Trappe, Ballesta, and Mazoyer}}]{cipelletti.l.03.time-resolved.s257}
\bibinfo{author}{\bibfnamefont{L.}~\bibnamefont{Cipelletti}},
  \bibinfo{author}{\bibfnamefont{H.}~\bibnamefont{Bissig}},
  \bibinfo{author}{\bibfnamefont{V.}~\bibnamefont{Trappe}},
  \bibinfo{author}{\bibfnamefont{P.}~\bibnamefont{Ballesta}}, \bibnamefont{and}
  \bibinfo{author}{\bibfnamefont{S.}~\bibnamefont{Mazoyer}},
  \bibinfo{journal}{J. Phys.: Condens. Matter} \textbf{\bibinfo{volume}{15}},
  \bibinfo{pages}{S257} (\bibinfo{year}{2003}).

\bibitem[{\citenamefont{Sibani and
  Jensen}(2005)}]{sibani.p.05.intermittency.563}
\bibinfo{author}{\bibfnamefont{P.}~\bibnamefont{Sibani}} \bibnamefont{and}
  \bibinfo{author}{\bibfnamefont{H.~J.} \bibnamefont{Jensen}},
  \bibinfo{journal}{Europhys. Lett.} \textbf{\bibinfo{volume}{69}},
  \bibinfo{pages}{563} (\bibinfo{year}{2005}).

\bibitem[{\citenamefont{Sibani}(2006{\natexlab{a}})}]{sibani.p.06.mesoscopic.6%
9}
\bibinfo{author}{\bibfnamefont{P.}~\bibnamefont{Sibani}},
  \bibinfo{journal}{Europhys. Lett.} \textbf{\bibinfo{volume}{73}},
  \bibinfo{pages}{69} (\bibinfo{year}{2006}{\natexlab{a}}).

\bibitem[{\citenamefont{Sibani}(2006{\natexlab{b}})}]{sibani.p.06.aging.031115}
\bibinfo{author}{\bibfnamefont{P.}~\bibnamefont{Sibani}},
  \bibinfo{journal}{Phys. Rev. E} \textbf{\bibinfo{volume}{74}},
  \bibinfo{pages}{031115/1} (\bibinfo{year}{2006}{\natexlab{b}}).

\bibitem[{\citenamefont{Lipowski and
  Johnston}(2000)}]{lipowski.a.00.cooling-rate.6375}
\bibinfo{author}{\bibfnamefont{A.}~\bibnamefont{Lipowski}} \bibnamefont{and}
  \bibinfo{author}{\bibfnamefont{D.~A.} \bibnamefont{Johnston}},
  \bibinfo{journal}{Phys. Rev. E} \textbf{\bibinfo{volume}{61}},
  \bibinfo{pages}{6375} (\bibinfo{year}{2000}).

\bibitem[{\citenamefont{Swift et~al.}(2000)\citenamefont{Swift, Bokil,
  Travasso, and Bray}}]{swift.r.00.glassy.11494}
\bibinfo{author}{\bibfnamefont{M.~R.} \bibnamefont{Swift}},
  \bibinfo{author}{\bibfnamefont{H.}~\bibnamefont{Bokil}},
  \bibinfo{author}{\bibfnamefont{R.~D.~M.} \bibnamefont{Travasso}},
  \bibnamefont{and} \bibinfo{author}{\bibfnamefont{A.~J.} \bibnamefont{Bray}},
  \bibinfo{journal}{Phys. Rev. B} \textbf{\bibinfo{volume}{62}},
  \bibinfo{pages}{11494} (\bibinfo{year}{2000}).

\bibitem[{\citenamefont{Sibani and Littlewood}(1993)}]{sibani.p.93.slow.1482}
\bibinfo{author}{\bibfnamefont{P.}~\bibnamefont{Sibani}} \bibnamefont{and}
  \bibinfo{author}{\bibfnamefont{P.~B.} \bibnamefont{Littlewood}},
  \bibinfo{journal}{Phys. Rev. Lett.} \textbf{\bibinfo{volume}{71}},
  \bibinfo{pages}{1482} (\bibinfo{year}{1993}).

\bibitem[{\citenamefont{Sibani and Dall}(2003)}]{sibani.p.03.log-poisson.8}
\bibinfo{author}{\bibfnamefont{P.}~\bibnamefont{Sibani}} \bibnamefont{and}
  \bibinfo{author}{\bibfnamefont{J.}~\bibnamefont{Dall}},
  \bibinfo{journal}{Europhys. Lett.} \textbf{\bibinfo{volume}{64}},
  \bibinfo{pages}{8} (\bibinfo{year}{2003}).

\bibitem[{\citenamefont{Vincent}(1991)}]{vincent.e.91.slow.209}
\bibinfo{author}{\bibfnamefont{E.}~\bibnamefont{Vincent}}, in
  \emph{\bibinfo{booktitle}{Recent Progress in Random Magnets}}, edited by
  \bibinfo{editor}{\bibfnamefont{D.~H.} \bibnamefont{Ryan}}
  (\bibinfo{publisher}{Mc Gill University}, \bibinfo{year}{1991}), pp.
  \bibinfo{pages}{209--246}.

\bibitem[{\citenamefont{Bouchaud and Dean}(1995)}]{bouchaud.j.95.aging.265}
\bibinfo{author}{\bibfnamefont{J.-P.} \bibnamefont{Bouchaud}} \bibnamefont{and}
  \bibinfo{author}{\bibfnamefont{D.~S.} \bibnamefont{Dean}},
  \bibinfo{journal}{J. Phys. I} \textbf{\bibinfo{volume}{5}},
  \bibinfo{pages}{265} (\bibinfo{year}{1995}).

\bibitem[{\citenamefont{Krawczyk and
  Kobe}(2002)}]{krawczyk.j.02.low-temperature.302}
\bibinfo{author}{\bibfnamefont{J.}~\bibnamefont{Krawczyk}} \bibnamefont{and}
  \bibinfo{author}{\bibfnamefont{S.}~\bibnamefont{Kobe}},
  \bibinfo{journal}{Physica A} \textbf{\bibinfo{volume}{315}},
  \bibinfo{pages}{302} (\bibinfo{year}{2002}).

\bibitem[{\citenamefont{Sibani and
  Hoffmann}(1989)}]{sibani.p.89.hierarchical.2853}
\bibinfo{author}{\bibfnamefont{P.}~\bibnamefont{Sibani}} \bibnamefont{and}
  \bibinfo{author}{\bibfnamefont{K.~H.} \bibnamefont{Hoffmann}},
  \bibinfo{journal}{Phys. Rev. Lett.} \textbf{\bibinfo{volume}{63}},
  \bibinfo{pages}{2853} (\bibinfo{year}{1989}).

\bibitem[{\citenamefont{Sibani et~al.}(1993)\citenamefont{Sibani, Sch\"on,
  Salamon, and Andersson}}]{sibani.p.93.emergent.479}
\bibinfo{author}{\bibfnamefont{P.}~\bibnamefont{Sibani}},
  \bibinfo{author}{\bibfnamefont{J.~C.} \bibnamefont{Sch\"on}},
  \bibinfo{author}{\bibfnamefont{P.}~\bibnamefont{Salamon}}, \bibnamefont{and}
  \bibinfo{author}{\bibfnamefont{J.-O.} \bibnamefont{Andersson}},
  \bibinfo{journal}{Europhys. Lett.} \textbf{\bibinfo{volume}{22}},
  \bibinfo{pages}{479} (\bibinfo{year}{1993}).

\bibitem[{\citenamefont{Hoffmann et~al.}(1997)\citenamefont{Hoffmann, Schubert,
  and Sibani}}]{hoffmann.k.97.age.613}
\bibinfo{author}{\bibfnamefont{K.~H.} \bibnamefont{Hoffmann}},
  \bibinfo{author}{\bibfnamefont{S.}~\bibnamefont{Schubert}}, \bibnamefont{and}
  \bibinfo{author}{\bibfnamefont{P.}~\bibnamefont{Sibani}},
  \bibinfo{journal}{Europhys. Lett.} \textbf{\bibinfo{volume}{38}},
  \bibinfo{pages}{613} (\bibinfo{year}{1997}).

\bibitem[{\citenamefont{Sibani and
  Hoffmann}(1991)}]{sibani.p.91.relaxation.423}
\bibinfo{author}{\bibfnamefont{P.}~\bibnamefont{Sibani}} \bibnamefont{and}
  \bibinfo{author}{\bibfnamefont{K.~H.} \bibnamefont{Hoffmann}},
  \bibinfo{journal}{Europhys. Lett.} \textbf{\bibinfo{volume}{16}},
  \bibinfo{pages}{423} (\bibinfo{year}{1991}).

\bibitem[{\citenamefont{Uhlig et~al.}(1995)\citenamefont{Uhlig, Hoffmann, and
  Sibani}}]{uhlig.c.95.relaxation.409}
\bibinfo{author}{\bibfnamefont{C.}~\bibnamefont{Uhlig}},
  \bibinfo{author}{\bibfnamefont{K.~H.} \bibnamefont{Hoffmann}},
  \bibnamefont{and} \bibinfo{author}{\bibfnamefont{P.}~\bibnamefont{Sibani}},
  \bibinfo{journal}{Z. Phys. B} \textbf{\bibinfo{volume}{96}},
  \bibinfo{pages}{409} (\bibinfo{year}{1995}).

\bibitem[{\citenamefont{Dall and Sibani}(2001)}]{dall.j.01.faster.260}
\bibinfo{author}{\bibfnamefont{J.}~\bibnamefont{Dall}} \bibnamefont{and}
  \bibinfo{author}{\bibfnamefont{P.}~\bibnamefont{Sibani}},
  \bibinfo{journal}{Comp. Phys. Comm.} \textbf{\bibinfo{volume}{141}},
  \bibinfo{pages}{260} (\bibinfo{year}{2001}).

\bibitem[{\citenamefont{Krug}(2007)}]{krug.j.07.records.P07001}
\bibinfo{author}{\bibfnamefont{J.}~\bibnamefont{Krug}},
  \bibinfo{journal}{Journal of Statistical Mechanics} p.
  \bibinfo{pages}{P07001} (\bibinfo{year}{2007}).

\bibitem[{\citenamefont{Hoffmann et~al.}(1985)\citenamefont{Hoffmann,
  Grossmann, and Wegner}}]{hoffmann.k.85.random.401}
\bibinfo{author}{\bibfnamefont{K.~H.} \bibnamefont{Hoffmann}},
  \bibinfo{author}{\bibfnamefont{S.}~\bibnamefont{Grossmann}},
  \bibnamefont{and} \bibinfo{author}{\bibfnamefont{F.~J.}
  \bibnamefont{Wegner}}, \bibinfo{journal}{Z. Phys. B}
  \textbf{\bibinfo{volume}{60}}, \bibinfo{pages}{401} (\bibinfo{year}{1985}).

\bibitem[{\citenamefont{Hoffmann and
  Sibani}(1988)}]{hoffmann.k.88.diffusion.4261}
\bibinfo{author}{\bibfnamefont{K.~H.} \bibnamefont{Hoffmann}} \bibnamefont{and}
  \bibinfo{author}{\bibfnamefont{P.}~\bibnamefont{Sibani}},
  \bibinfo{journal}{Phys. Rev. A} \textbf{\bibinfo{volume}{38}},
  \bibinfo{pages}{4261} (\bibinfo{year}{1988}).

\bibitem[{\citenamefont{Ball et~al.}(1996)\citenamefont{Ball, Berry, Kunz, Li,
  Proykova, and Wales}}]{ball.k.96.from.963}
\bibinfo{author}{\bibfnamefont{K.~D.} \bibnamefont{Ball}},
  \bibinfo{author}{\bibfnamefont{R.~S.} \bibnamefont{Berry}},
  \bibinfo{author}{\bibfnamefont{R.~E.} \bibnamefont{Kunz}},
  \bibinfo{author}{\bibfnamefont{F.-Y.} \bibnamefont{Li}},
  \bibinfo{author}{\bibfnamefont{A.}~\bibnamefont{Proykova}}, \bibnamefont{and}
  \bibinfo{author}{\bibfnamefont{D.~J.} \bibnamefont{Wales}},
  \bibinfo{journal}{Science} \textbf{\bibinfo{volume}{271}},
  \bibinfo{pages}{963} (\bibinfo{year}{1996}).

\bibitem[{\citenamefont{Geppert et~al.}(1997)\citenamefont{Geppert, Rieger, and
  Schreckenberg}}]{geppert.u.97.hierarchical.l393}
\bibinfo{author}{\bibfnamefont{U.}~\bibnamefont{Geppert}},
  \bibinfo{author}{\bibfnamefont{H.}~\bibnamefont{Rieger}}, \bibnamefont{and}
  \bibinfo{author}{\bibfnamefont{M.}~\bibnamefont{Schreckenberg}},
  \bibinfo{journal}{J. Phys. A: Math. Gen.} \textbf{\bibinfo{volume}{30}},
  \bibinfo{pages}{L393} (\bibinfo{year}{1997}).

\bibitem[{\citenamefont{Hoffmann}(1999)}]{hoffmann.k.99.slow.30}
\bibinfo{author}{\bibfnamefont{K.~H.} \bibnamefont{Hoffmann}},
  \bibinfo{journal}{Comp. Phys. Comm.} \textbf{\bibinfo{volume}{121-122}},
  \bibinfo{pages}{30} (\bibinfo{year}{1999}).

\bibitem[{\citenamefont{de~Oliveira et~al.}(1999)\citenamefont{de~Oliveira,
  Fontanari, and Stadler}}]{oliveira.v.99.metastable.8793}
\bibinfo{author}{\bibfnamefont{V.~M.} \bibnamefont{de~Oliveira}},
  \bibinfo{author}{\bibfnamefont{J.~F.} \bibnamefont{Fontanari}},
  \bibnamefont{and} \bibinfo{author}{\bibfnamefont{P.~F.}
  \bibnamefont{Stadler}}, \bibinfo{journal}{J. Phys. A: Math. Gen.}
  \textbf{\bibinfo{volume}{32}}, \bibinfo{pages}{8793} (\bibinfo{year}{1999}).

\bibitem[{\citenamefont{Hordijk et~al.}(2003)\citenamefont{Hordijk, Fontanari,
  and Stadler}}]{hordijk.w.03.shapes.3671}
\bibinfo{author}{\bibfnamefont{W.}~\bibnamefont{Hordijk}},
  \bibinfo{author}{\bibfnamefont{J.~F.} \bibnamefont{Fontanari}},
  \bibnamefont{and} \bibinfo{author}{\bibfnamefont{P.~F.}
  \bibnamefont{Stadler}}, \bibinfo{journal}{J. Phys. A: Math. Gen.}
  \textbf{\bibinfo{volume}{36}}, \bibinfo{pages}{3671} (\bibinfo{year}{2003}).

\bibitem[{\citenamefont{Wales}(2003)}]{wales.d.03.energy.book}
\bibinfo{author}{\bibfnamefont{D.~J.} \bibnamefont{Wales}},
  \emph{\bibinfo{title}{Energy landscapes}}, Cambridge molecular science
  (\bibinfo{publisher}{Cambridge Univ. Press}, \bibinfo{address}{Cambridge,
  UK}, \bibinfo{year}{2003}).

\bibitem[{\citenamefont{Schubert and Hoffmann}(2004)}]{schubert.s.04.aging.118}
\bibinfo{author}{\bibfnamefont{S.}~\bibnamefont{Schubert}} \bibnamefont{and}
  \bibinfo{author}{\bibfnamefont{K.~H.} \bibnamefont{Hoffmann}},
  \bibinfo{journal}{Europhys. Lett.} \textbf{\bibinfo{volume}{66}},
  \bibinfo{pages}{118} (\bibinfo{year}{2004}).

\bibitem[{\citenamefont{Schubert and
  Hoffmann}(2006)}]{schubert.s.06.structure.191}
\bibinfo{author}{\bibfnamefont{S.}~\bibnamefont{Schubert}} \bibnamefont{and}
  \bibinfo{author}{\bibfnamefont{K.~H.} \bibnamefont{Hoffmann}},
  \bibinfo{journal}{Comp. Phys. Comm.} \textbf{\bibinfo{volume}{174}},
  \bibinfo{pages}{191} (\bibinfo{year}{2006}).

\bibitem[{\citenamefont{Schreckenberg}(1985)}]{schreckenberg.m.85.long.483}
\bibinfo{author}{\bibfnamefont{M.}~\bibnamefont{Schreckenberg}},
  \bibinfo{journal}{Z. Phys. B} \textbf{\bibinfo{volume}{60}},
  \bibinfo{pages}{483} (\bibinfo{year}{1985}).

\bibitem[{\citenamefont{Dall and Sibani}(2003)}]{dall.j.03.exploring.233}
\bibinfo{author}{\bibfnamefont{J.}~\bibnamefont{Dall}} \bibnamefont{and}
  \bibinfo{author}{\bibfnamefont{P.}~\bibnamefont{Sibani}},
  \bibinfo{journal}{Eur. Phys. J. B} \textbf{\bibinfo{volume}{36}},
  \bibinfo{pages}{233} (\bibinfo{year}{2003}).

\bibitem[{\citenamefont{Boettcher and
  Sibani}(2005)}]{boettcher.s.05.comparing.317}
\bibinfo{author}{\bibfnamefont{S.}~\bibnamefont{Boettcher}} \bibnamefont{and}
  \bibinfo{author}{\bibfnamefont{P.}~\bibnamefont{Sibani}},
  \bibinfo{journal}{Eur. Phys. J. B} \textbf{\bibinfo{volume}{44}},
  \bibinfo{pages}{317} (\bibinfo{year}{2005}).

\end{thebibliography}
\bibliographystyle{apsrev}
\end{document}